\documentstyle[12pt]{article}
\begin{document}
\thispagestyle{empty}
\begin{raggedleft}
UR-1410\\
ER-40685-857\\
hep-th/9502056\\
Dec.\ 1994\\
\end{raggedleft}
$\phantom{x}$\vskip 0.618cm\par
{\huge \begin{center}ON THE COVARIANTIZATION OF THE CHIRAL CONSTRAINTS
\footnote{This work is
supported by CNPq, Bras\'\i lia, Brasil}
\end{center}}\par
\begin{center}
$\phantom{X}$\\
{\Large E.M.C. de Abreu and  C.Neves }\\[3ex]
{\em Instituto de F\'\i sica\\
Universidade Federal do Rio de Janeiro\\
Rio de Janeiro\\
Brasil\\}
\end{center}\par
\begin{center}
$\phantom{X}$\\
{\Large Clovis Wotzasek\footnote{Permanent address:
Instituto de F\'\i sica,
Universidade Federal do Rio de Janeiro, Brasil}}\\[3ex]
{\em Department of Physics and Astronomy\\
University of Rochester\\
Rochester, NY 14627 \\
USA}
\end{center}\par
\begin{abstract}

\noindent  We show that a complete covariantization of the chiral constraint in
the Floreanini-Jackiw necessitates an infinite number of auxiliary Wess-Zumino
fields otherwise the covariantization is only partial and unable to remove the
nonlocality in the chiral boson operator.  We comment on recent
works that claim to obtain covariantization through the use of
Batalin-Fradkin-Tyutin method, that uses just one Wess-Zumino field.

\end{abstract}
\vfill
\newpage

The quantization of chiral boson in two-dimensions is a very
interesting theoretical problem, which has appeared originaly in the
investigation of heterotic string \cite{string}, and became quite
important in
the study of fractional quantum Hall efect \cite{hall}.  This problem has a
simple solution in the Hamiltonian language while it has been beset
with enormous difficulties in the Lagrangian side.  One of these
problems is the covariantization of the second-class chiral
constraint, i.e., the transformation from second to first-class,
which is the object of investigation in this paper.

The usual Lagrangian route to chiral bosonization starts with
a scalar field and projects out
one of the chiral components by means of the chiral constraint
$\partial_{\pm}\phi\approx0$.  According to Dirac's theory of constrained
systems \cite{dirac}, this is a second-
class constraint, but in order to avoid the Lagrange multiplier to
become dynamical, one needs it to be first-class.  There has been two
main routes to the covariantization of the chiral constraint.
Siegel \cite{siegel}
proposed to set to zero one component of the energy-momentum tensor resulting
an action with the chiral constraint squared, which has a
reparametrization invariance called as Siegel symmetry.  In the quantum level,
however, Siegel symmetry becomes second-class again on account of
the central extension of the conformal algebra of the energy-momentum
tensor \cite{conformal}.
A simple solution to the anomaly problem was given by Hull with
the introduction of a new set of
auxiliary fields called as no-movers \cite{nomovers}.  Besides the
anomaly problem, to produce first-class constraints by squaring
second-class constraints has been criticized \cite{critica}.  This sort of
primary constraint does not produce the complete set of constraints
when Dirac's
algorithm is employed, and is (infinitely) reducible.

The second route to covariantization starts with the
Floreanini-Jackiw model\cite{fj} which is a singular theory from Dirac's
point of view, the resulting constraint being the second-class chiral
constraint.  The constraint's nature is changed \`a la
Faddeev-Shatashvili\cite{fad-shatash} with the introduction of Wess-Zumino
auxiliary fields.  The covariantization of the chiral constraint
in this route has been proposed in two
conceptually differents papers: in Ref.\cite{mwy}, an infinite family
of scalar
fields, coupled by a combination of right and left chiral constraints
carefully adjusted to be first-class from the start, was shown to have a
single chiral boson in the spectrum by use of very elegant group
theoretical methods.
In Ref.\cite{clovis-prl}, the FJ chiral boson was iteratively
changed to modify the nature of chiral constraint to render it first-class.
Following this route the modified Floreanini-Jackiw model with the original
chiral field plus the set of Wess-Zumino fields will always have one left over
second-class constraint that must as well be converted into first-class if a
complete covariantization is desired. It is worth of mention that, as
explained by Boyanovsky\cite{boya}, the inclusion of constraint conversion
terms do not change the physical spectrum of the theory but only the
unphysical
sector  of the Hilbert space.  Therefore the chiral boson spectrum which we
begin with will not be changed during the constraint conversion process.
For the case of FJ-chiral boson, the
algorithm in \cite{clovis-prl} revealed the necessity of an infinite
number of auxiliary fields, thus stablishing a connection with the
model in \cite{mwy}.  This necessity of an infinite set of auxiliary fields
in the the covariantization process of the chiral boson operator can be
traced back to the fact that the operator of the chiral boson action
is the square root of the
D'Lambertian operator, therefore of nonlocal nature.  Consequently, in
following this route we are basically exchanching the non-locality by
auxiliary fields, as we are used to. The equivalence of proposals
\cite{mwy} and \cite{clovis-prl} has been proved in \cite{clovis-equiv}.  We
mention that the use of infinite auxiliary fields in the covariantization
process of second-class constraints has also been used by Mikovic et al
\cite{mikovic} in the context of the super-particle quantization.

It is our intention in this paper to work out the possibility of stopping the
constraint conversion process, mentioned above, at any (finite) stage, say
after N iterations, even if a second-class constraint remains, and
study its consequences. It is
expected that in doing this we would destroy the possibility of obtaining
the complete Wess-Zumino Lagrangian, and the theory would only be partially
covariantized.  Furthermore, one should not expect to completely remove
the non-locality of the chiral boson operator by stopping the constraint
conversion process at any finite N, i.e., with use of a finite number of
auxiliary fields. We will see that these expectations are indeed fullfilled and
in doing so we get a theory that is neither fully covariant nor local.
We plan to stop the constraint conversion process in
two different ways and verify their equivalence. Firstly we will follow a
constraint conversion method introduced by one of us some years ago in the
context of the chiral Schwinger model \cite{clovis-ijmpa}. As
explained there, this method is unable to provide
the complete Wess-Zumino Lagrangian,
but corresponds only to some gauge-fixed version of it.  Putting in
another words, this procceedure only restores part of the symmetry
lost during the quantization process, and this happens due to the
existence of a ``stopping rule'' that stops the covariantization process
even if a second-class constraint is still present (see below).  Another way
of doing it is by a simple elimination of the left over second-class constraint
using Dirac brackets for the variables involved.  This can be done after
an arbitrary N number of steps.  It will then be clear that the results of
the first method corresponds to choose to stop at N=1 in the second method.  At
this point we make contact with a third proposal for the covariant
quantization of FJ chiral boson which has recently appeared in the
literature\cite{ghosh1,ghosh2} (see also \cite{amorim}).  Differently of
the other two schemes, these works claim to make use of a single
Wess-Zumino auxiliary
field to produce the constraint conversion into first-class.  In
\cite{ghosh1,ghosh2}, use is made of the constraint conversion technique
proposed by Fradkin and colaborators \cite{fbt}.  The basic idea in
\cite{fbt} is to produce the change in the constraint's nature,
by enlarging the Hamiltonian phase-space with ghosts and Lagrange
multipliers of opposite statistics in order to obtain the nilpotency
of the BRST charge.

We show that the results we found using both methods discussed above are
equivalent to \cite{ghosh1,ghosh2} in the sense that the
covariantization of the initial chiral constraint is acchieved with only
one auxiliar field.  From these results we conclude that this third proposal
for covariantization with only one
Wess-Zumino field is only able to acchive partial covariantization, and is
equivalent to stopping the
iterative method of \cite{clovis-prl} at the N=1 stage.

To begin with, let us review the basic idea proposed in
\cite{clovis-ijmpa}.  Suppose a non-invariant theory, described by 2n
phase-space variables, a Hamiltonian $H_0$, and a set of 2m
second-class constraints $\Omega_a$.  The Dirac matrix of the
constraints is defined as

\begin{equation}
\label{dirac-matrix}
C_{ab} = \{\Omega_a,\Omega_b\}
\end{equation}

\noindent and satisfy $det\,C_{ab}\neq0$ due to the second-class
nature of the constraints.
Call $H
= H_0 + \lambda_a\Omega_a$ the total or Dirac Hamiltonian. Since there
is no avaiable symmetry, by assumption, the Lagrange multipliers
$\lambda_a$ are not arbitrary and can be determined from the
consistency on the time development of $\Omega_a$.  Next, we introduce
the auxiliary Wess-Zumino fields $\theta_a$, which are choosen to satisfy

\begin{equation}
\label{theta}
\{\theta_a,\theta_b\} = C^{-1}_{ab}
\end{equation}

\noindent It is then easy to verify that the new set of constraints
$\tilde{\Omega}_a$, defined by

\begin{equation}
\label{new-constraint}
\Omega_a\rightarrow\tilde{\Omega}_a=\Omega_a+\sum_bC_{ab}\,\theta_b
\end{equation}

\noindent is first-class $\{\tilde{\Omega}_a,\tilde{\Omega}_b\}=0$.
The new
set of constraints may, however, necessitates a secondary
constraint manifold to guarantee its time development stability,
and it is possible that some of these secondary constraints
may be second-class again.  In this eventuality a new set of
Wess-Zumino fields would become necessary and so on.  In order to
avoid such behavior, we introduce a ``stopping rule'' by demanding the
first-class set of constraints (\ref{new-constraint}) to commute
with a generalized
Hamiltonian $\tilde{H}$, obtained from the original one, as a (functional)
Taylor series of the Wess-Zumino fields

\begin{equation}
\label{new-hamiltonian}
H\rightarrow\tilde{H}=H_0+\sum_aM_a\theta_a+
\frac{1}{2}\sum_{ab}M_{ab}\theta_a\theta_b+\ldots
\end{equation}

\noindent The ``stopping rule'' mentioned above takes the form

\begin{equation}
\label{stopping-rule}
\{\tilde{\Omega_a},\tilde{H}\}=0
\end{equation}

\noindent Solving equation (\ref{stopping-rule}), order by order
in $\theta$, fixes the expansion coeficient as

\begin{eqnarray}
\label{coefficients}
& M^{(1)}_a=\{H_0,\Omega_a\} \nonumber \\
\ \\
& M^{(2)}_{ab}=\{M_a^{(1)},\Omega_b\} \nonumber
\end{eqnarray}

\noindent and so on, as explained in \cite{clovis-ijmpa}.

Let us consider now the case of Floreanini-Jackiw chiral boson.
The Floreanini-Jackiw \cite{fj} chiral
boson theory is a first order model with the spectrum being
derived directly from its equation of motion, without the necessity of
external constraints.  The drawback of this formulation is
the lack of explicit Lorentz covariance, which makes the coupling with
gauge or gravitation fields rather difficult
\cite{sonnenschein,bellucci,bastianelli,harada}.
For the case of
the (left mover) FJ chiral  boson, described by

\begin{equation}
\label{florjack}
L=\int\,dx\,(\dot{\phi}\phi'-\phi'^{2})
\end{equation}

\noindent the Dirac Hamiltonian reads

\begin{eqnarray}
\label{dirac-hamiltonian}
& H=\int\,dx\,\left[\phi'^2+\lambda(\pi-\phi')\right] \nonumber \\
& =\int\,dx\,\pi\phi'
\end{eqnarray}

\noindent where $\pi=\delta{\cal L}\,/\,\delta\dot{\phi}$ is the canonical
momentum of $\phi(x)$ and $\lambda=\phi'$ has been determined from the
consistency condition for the evolution of the chiral constraint
$\Omega=\pi-\phi'$ .  The modified (first-class) constraint is,
according to definition (\ref{new-constraint}), given by

\begin{equation}
\label{1st-modified}
\Omega\rightarrow\tilde{\Omega}=\pi-\phi'-2\theta'
\end{equation}

\noindent where $\theta$ is the Wess-Zumino field, satisfying
condition (\ref{theta})

\begin{equation}
\label{theta-fj}
\{\theta'(x),\theta'(y)\}=\frac{1}{2}\,\delta'(x-y)
\end{equation}

\noindent which shows that this auxiliary field is itself of chiral
nature.  The other brackets are canonical

\begin{eqnarray}
\label{brackets}
\{\phi(x), \pi(y)\} &=& \delta(x-y)\nonumber\\
\{\phi(x),\theta(y)\}&=&0\nonumber\\
\{\pi(x),\theta(y)\}&=&0
\end{eqnarray}

\noindent To compute $\tilde{H}$ is now a pretty simple algebraic work.
Using (\ref{coefficients}) we obtain

\begin{eqnarray}
\label{coefficients-fj}
& M^{(1)}(x)= - 2\phi''(x) \nonumber \\
\ \\
& M^{(2)}(x,y)=-2\,\delta''(x-y) \nonumber
\end{eqnarray}

\noindent while all the remaing coefficients vanish.  Then,
using (\ref{coefficients-fj}) into (\ref{dirac-hamiltonian}) we get

\begin{equation}
\label{hamiltonian-fj}
\tilde{H}=\int\,dx\,(\phi'^2+2\phi'\theta'+\theta'^2)
=\int\,dx\,(\phi'+\theta')^2
\end{equation}

\noindent Notice that making use of (\ref{theta-fj}) and
(\ref{brackets}) we obtain a first-class algebra for the (modified)
chiral constraint

\begin{equation}
\{\tilde{\Omega}(x), \tilde{\Omega}(y)\}=0
\end{equation}

\noindent We have thus succeeded in turning into first-class the
original second-class constraint of the Floreanini-Jackiw model.
To make contact with the results in Ref. \cite{ghosh1,ghosh2},
we observe
that a concrete realization the algebra (\ref{theta-fj}) in terms of a
canonically conjugated pair is given by

\begin{equation}
\label{new-pair}
\theta'=\frac{1}{2}\xi+\frac{1}{2}\eta'
\end{equation}

\noindent with

\begin{equation}
\{\eta(x),\xi(y)\}=\delta(x-y)
\end{equation}

\noindent In this representation, the canonical Hamiltonian
(\ref{hamiltonian-fj}) and the first-class chiral constraint become

\begin{eqnarray}
\label{ghosh-result1}
& \tilde{H}=\int\,dx\,(\phi'+\frac{1}{2}\xi+\frac{1}{2}\eta')^2 \\
\label{ghosh-result2}
& \tilde{\Omega}= \pi-\phi'-\xi-\eta'
\end{eqnarray}

\noindent which are the results of  \cite{ghosh1,ghosh2}.  Quite clearly this
first-class chiral constraint (\ref{ghosh-result2})
is a combination of
a right and a left chiral constraint

\begin{equation}
\label{ghosh-constraint}
\tilde{\Omega}=\Omega^{(-)}-\Omega^{(+)}
\end{equation}

\noindent where the chiral constraints $\Omega^{(\pm)}(x)$ satisfy two
uncoupled Kac-Moody algebras

\begin{eqnarray}
\label{kac-moody}
& \left\{\Omega^{(\pm)}(x),\Omega^{(\pm)}(y)\right\}
=(\mp)2\delta'(x-y) \nonumber \\
\ \\
& \left\{\Omega^{(+)}(x),\Omega^{(-)}(y)\right\}=0 \nonumber
\end{eqnarray}

\noindent Notice that the constraint (\ref{ghosh-result2}) is of
the same sort as that introduced in Refs. \cite{mwy} and \cite{clovis-prl}
showing that some connection might exist between these results.
It is precisely this relationship that we intent to clarify below.

In the sequence we want to show that the above result,
Eqs.(\ref{1st-modified}) and
(\ref{hamiltonian-fj}) (or (\ref{ghosh-result1})
and (\ref{ghosh-result2})) can be obtained as a
particularization of the
method given in Ref. \cite{clovis-prl} if, after one iteration, the emergent
second-class constraint instead of being covariantized is
strongly resolved \`a la Dirac.  To this end we quickly review the basic
points of \cite{clovis-prl}, taking the
opportunity to present a new approach to that problem.  Indeed, the
results of \cite{clovis-prl} after N iterations,
are easily found if in the FJ Lagrangian (\ref{florjack})
we make the substitution

\begin{equation}
\label{substitution}
\phi_0\rightarrow\phi_0+\phi_1+ \ldots +\phi_N
\end{equation}

\noindent where $\phi_0$ represents the original chiral boson field
($\phi \rightarrow \phi_0$) and the remaining $\phi_k$
are N auxiliary Wess-Zumino fields.  Introducing (\ref{substitution})
into (\ref{florjack}) gives

\begin{equation}
\label{nth-iterated}
L^{(0)}\rightarrow L^{(N)}=\sum_{k=0}^{N}\int\,dx\,(\dot{\phi_k}\phi_k'
-\phi_k'^2)+
\sum_{m=k+1}^{N}\,\sum_{k=0}^{N-1}\int\,dx\,(\dot{\phi_k}-\phi_k')\phi_m'
\end{equation}

\noindent which is the result in \cite{clovis-prl}.  Consequently, the original
chiral constraint $\Omega_0$ present in the Floreanini-Jackiw action
is  transformed into a set of $(N+1)$ constraints

\begin{eqnarray}
\label{iterated-constraints}
& \Omega_0\rightarrow\{\Omega_0,\Omega_1, \dots ,\Omega_{N}\} \nonumber \\
\ \\
& \Omega_k\approx0 \nonumber
\end{eqnarray}

\noindent Explicitly, these constraints read

\begin{eqnarray}
\label{prl-constraints}
& \Omega_N=\pi_N-\phi_N' \nonumber \\
\ \\
& \Omega_k=\pi_k-\phi_k'-2\sum_{m=k+1}^{N}\phi_m' \nonumber
\end{eqnarray}

\noindent and the canonical Hamiltonian becomes

\begin{equation}
H\rightarrow H_N=\int\,dx\sum_{m,n=0}^{N}\phi_m'\phi_n'
\end{equation}

\noindent which is easily obtained from ${\cal H}_0=\phi'^2$ using the
substitution (\ref{substitution}).  As discussed above,
the last constraint $\Omega_N$ always
break the emergent symmetry by keeping the second-class character of
the iterated theory.

Suppose we decide to stop the iteration at this point.  This can be
done by explicitly solving the left over second-class constraint
$\Omega_N(x)\approx0$.  This is possible as long as we use the
corresponding Dirac bracket for the variables involved.  The relevant Dirac
bracket in this case is, from (\ref{prl-constraints})

\begin{equation}
\{\phi_N,\phi_N\}^*=\frac{1}{2}\,\delta'(x-y)
\end{equation}

\noindent and the remaining variables satisfy

\begin{equation}
\{\phi_k,\phi_m\}^*=0
\end{equation}

\noindent In particular, let us consider the partial covariantization
scheme above for the case $N=1$.  We have then

\begin{equation}
\{\phi_1,\phi_1\}^*=\frac{1}{2}\delta'(x-y)
\end{equation}

\noindent and

\begin{equation}
\{\phi_0,\phi_0\}^*=0
\end{equation}

\noindent The canonical Hamiltonian reads

\begin{equation}
H_1=\int\,dx(\phi_0'-\phi_1')^2
\end{equation}

\noindent and the chiral constraint becomes

\begin{equation}
\Omega_0=\pi_0 - \phi'_0 - 2\phi'_1
\end{equation}

\noindent If we relabel the fields as $\phi_0\rightarrow\phi$ and
$\phi_1\rightarrow\theta$, we reobtain the results above
(Eqs.(\ref{1st-modified}) and (\ref{hamiltonian-fj})) therefore
justifying our claiming that those schemes are only
able to provide a partial covariantization for the chiral boson theory.

In summary, we have worked the problem of covariantization of the
bosonic chiral constraint following the methodology of
Ref.\cite{clovis-ijmpa}.  Differently from the approach of
\cite{clovis-prl} where an infinite number of auxiliary fields is
necessary, this one only needs one Wess-Zumino field.  This result
was shown to agree with with a paper by Ghosh\cite{ghosh1,ghosh2} using
\cite{fbt}.  We have then showed that the
covariantization with just one Wess-Zumino auxiliary field is
incomplete and corresponds to stopping the iterative process of
\cite{clovis-prl} after a single iteration, through the (strong)
resolution of the left over chiral constraint \`a la Dirac.  This
seem to be an expected result since the nonlocality of the chiral
boson operator points for the necessity of an infinite set of
auxiliary fields as found in \cite{clovis-prl}.  It is not clear for
us at the present if this results are indicating an essential
difficulty in the formalism of Ref.\cite{clovis-ijmpa} and \cite{fbt}
or are just accidental and restricted to this (uncommom) model. Work
in this direction is in progress.


\begin{thebibliography}{99}
\bibitem{string}D.J.Gross, J.A.Harvey, E.Martinec, R.Rhom,
Phys.Rev.Lett.54 (1985)502.
\bibitem{hall}See, for instance, "Quantum Hall Effect", M.Stone ed.
(World Scientific 1992).
\bibitem{dirac}P.A.M.Dirac, Lectures on Quantum
Mechanics (Yeshiva University Press, New York, N.Y.) 1964.
\bibitem{siegel}W.Siegel, Nucl.Phys.B238(1984)307.
\bibitem{conformal}R.Nepomechie , L.Mezincescu, Phys.Rev.D37 (1988) 3067
and A.Schwimmer , C. Imbimbo, Phys. Lett.B193 (1987) 455.
\bibitem{nomovers}C.Hull, Phys.Lett.B206(1988)234.
\bibitem{critica}B.McClain and Y.S.Wu, University of Utah preprint/1988
and M.Lled\'o and A.Restuccia, USB preprint/1990.
\bibitem{fj}R.Floreanini, R.Jackiw, Phys.Rev.Lett.59 (1987) 1873.
\bibitem{fad-shatash}L.D.Faddeev and L.Shatashvili,
Phys.Lett.B167(1986)225.
\bibitem{mwy}B.McClain, Y.S.Wu, F.Yu, Nucl.Phys.B343(1990)689.
\bibitem{clovis-prl}C.Wotzasek, Phys.Rev.Lett.{\bf 66}(1991)129-133
\bibitem{boya}D. Boyanovsky, J. Schmidt and M. F. L. Golterman,
Ann. Phys. (NY) 185 (1988) 111.
\bibitem{clovis-equiv}C.Wotzasek, Phys. Rev. D44 (1991) R1623.
\bibitem{mikovic}A.Mikovic, M.Rocek, W.Siegel, P. van Nieuvenhuizen, J.Yamron,
and A.E. van de Ven, Phys. Lett. B235 (1990) 106.
\bibitem{clovis-ijmpa}C.Wotzasek, Intl.J.Mod.Phys{\bf A5}(1990)1123.
\bibitem{ghosh1}Subir Ghosh, Phys. Rev. D49 (1994) 2990.
\bibitem{ghosh2}Subir Ghosh, Mod. Phys. Lett. A9 (1994) 535.
\bibitem{fbt}I.A.Batalin and E.S.Fradkin, Nucl.Phys.B279(1987)514 and
I.A.Batalin and I.V.Tyutin, Intl.J.Mod.Phys.A6(1991)3255.
\bibitem{sonnenschein}J.Sonenschein, Nucl.Phys.B309(1988)752.
\bibitem{bellucci}S. Bellucci, R. Brooks and J. Sonnenschein,
Nucl. Phys. B304 (1988) 173 and S. Bellucci, M. F. L. Golterman and D. N.
Petcher, Nucl. Phys. B326 (1989) 307.
\bibitem{bastianelli}F.Bastianelli and P.Van Nieuwenhuizen, Phys.Lett.B217
(1989) 98.
\bibitem{harada}K.Harada,Phys.Rev.Lett.64(1990)139.
\bibitem{amorim}After the completion of this work we became aware of the
paper by R.Amorim and J.Barcelos-Neto, ``BFT quantization of the FJ
chiral boson'', IF-UFRJ-13/94, on the covariantization of the chiral
constraint using the Batalin-Fradkin and Tyutin technique for constraint
conversion.  In this paper they were able to show that the Lagrangian for
the converted theory results being nonlocal, as expected.  We thank
R.Amorim for making a copy of that work avaiable for us prior to its
publication.
\end{thebibliography}
\end{document}